\documentclass[twocolumn,amsmath,amssymb,longbibliography]{revtex4-2}

\usepackage[utf8]{inputenc}
\usepackage[english]{babel}
\usepackage[dvipsnames]{xcolor}
\usepackage{graphicx}
\usepackage{hyperref}

\begin{document}
	
\title{High-resolution coincidence counting system\\
	for large-scale photonics applications}

\author{Josef Hlou\v{s}ek}
\email{hlousek@optics.upol.cz}
\author{Jan Grygar}
\author{Michal Dudka}
\author{Miroslav Je\v zek}
\email{jezek@optics.upol.cz}
\affiliation{Department of Optics, Faculty of Science, Palack\'y University,
             17.\ listopadu 12, 77900 Olomouc, Czech Republic}

\date{\today}

\begin{abstract}
The increasing complexity of the recent photonic experiments challenges developing efficient multi-channel coincidence counting systems with high-level functionality. Here, we report a coincidence unit able to count detection events ranging from singles to 16-fold coincidences with full channel-number resolution. The device operates within sub-100~ps coincidence time windows, with a maximum input frequency of 1.5~GHz and an overall jitter of less than 10~ps. The unit high-level timing performance renders it suitable for quantum photonic experiments employing low-timing-jitter single-photon detectors. Additionally, the unit can be used in complex photonic systems to drive feed-forward loops. We have demonstrated the developed coincidence counting unit in photon-number-resolving detection to directly quantify the statistical properties of light, specifically coherent and thermal states, with a fidelity exceeding 0.999 up to 60~photons.
\end{abstract}

\maketitle

\section{Introduction}
%%%CCU applications
A coincidence counting unit (CCU) is an essential tool widely employed in all applications requiring the detection of a large number of photons (or other particles) and processing the detected signals. Modern quantum experiments employ increasingly complex systems with a growing number of input and output channels \cite{Wang2018, Wang_2019, Zhong_2020}.
These large-scale photonics systems hinge on the ability to generate, control, and analyze the multi-photon quantum states \cite{Yao2012, Wang2016, Wang2018, Paesani2019} frequently used in quantum communications \cite{Aaronson2014, Pirandola2020}, quantum computation and simulations \cite{AspuruGuzik2012}. Particularly, complex coincidence processing has become an integral part of measuring unknown optical states by photon-number-resolving detectors based on multiplexing \cite{Jex1996,Vogel2012,Hlousek2019}. Detected statistical properties of light are routinely applied to quantify the non-classicality and quantum-non-Gaussianity \cite{Mandel1977, Saleh2016, Sperling2017, Straka2018, Bohmann2019}. Furthermore, the advanced functionality of these devices is of considerable interest in on-the-fly multifold coincidences analysis to control large-scale quantum systems via the feed-forward operation and photon-number-resolving post-selection \cite{Jezek2019, Walther2021}.

%%%CCU approaches
Conventional approaches to detect coincidences are 1. time to amplitude converter (TAC) together with a single or multi-channel analyzer, 2. time to digital converter (TDC) followed by postprocessing, and 3. overlap logic coincidence systems realized with discrete components or using a field-programmable gate array (FPGA). TAC and TDC both typically offer tens of picosecond resolution. TACs are not easily scaled up for multi-coincidence systems and possess a considerable dead time limiting rate throughput to tens of thousands of events per second \cite{Beck2007,Beck2009}.
TDC-based solutions stream time tags to a computer for further processing. Therefore, a large amount of data is processed offline. Alternatively, TDCs are combined with an FPGA for the subsequent processing \cite{Wahl2013}. Pulse overlap coincidence systems use fast logic gates and multiplexers to capture detection events and detect coincidences \cite{Beck2009}. Coincidence counting and histogramming could be programmed into a microcontroller \cite{Weinfurter2005} or FPGA \cite{Beck2011}. Functional blocks such as internal delay lines, coincidence counters, and a processor can all be integrated within a single FPGA chip \cite{Park2015,Jiang2016,Gupta2018,Jiang2018}. Lately, multi-channel TDC-based coincidence counter architecture in the same FPGA chip was introduced \cite{Arabul2020}.

%%%the presented CCU
In this paper, we report an ultra-fast electronic multi-channel CCU, producing a histogram of all possible coincidence events for up to 16 constituent detectors. The device performs a real-time classification of all possible detection events in a $2^{16}$- element histogram with the rate of up to 3 million events per second. We have conducted a comprehensive characterization of the presented CCU, revealing excellent performance parameters, including a sub-100~ps coincidence window, sub-10~ps jitter, and an ultra-low coincidence error probability. Furthermore, we demonstrated the CCU versatility by implementing the multi-photon counting experiment to fully characterize the statistical properties of incident light.

\section{Coincidence counting unit}
\begin{figure}[t!] \centering
	\includegraphics[width=1.0 \columnwidth]{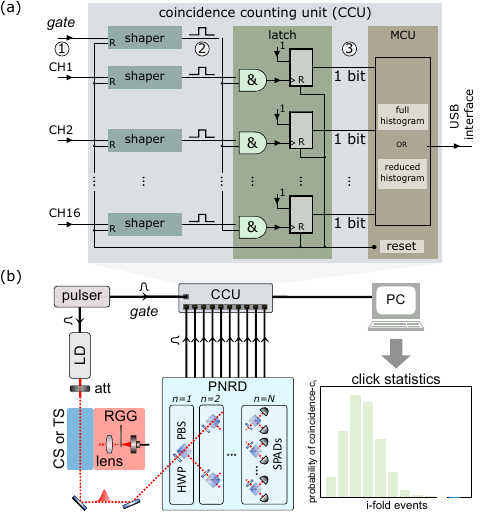}
	\caption{
		(a) A block diagram of the coincidence counting unit (CCU): the shapers prepare square pulses; coincidences are evaluated as an overlap (\&) of the pulses and can be directly used or counted and sorted in histograms by the microcontroller unit (MCU). (b) For photon-number-resolved detection (PNRD), the CCU is connected to a spatial-multiplexed optical network consisting of tunable beam splitters and single-photon avalanche diodes (SPADs). The resulting coincidence histogram allows for statistical analysis of the incident signal.
	}
	\label{fig_1}
\end{figure}

%%%Device design and operation
Figure~\ref{fig_1}(a) shows an overview of the CCU architecture based on fast positive emitter-coupled logic (ECL) offering high-resolution coincidence counting. The CCU consists of input signal overlap logic and data processing unit. The device accepts 16 data inputs and a single gate input and yields the complete histogram of $2^{16}$ multi-coincidences of the inputs within the gate signal. Each input channel contains a shaping circuit with high-resolution programmable delay lines to detect an input signal edge and provides the output ECL pulse of a given width and delay. The pulse width corresponds to half of the coincidence window, which can be tuned independently for each channel. After the signal shaping, the signal and gate pulse enable inputs of ECL latch circuitry. The latch records the overlap of individual signal pulses and the gate signal and stores information about coincidence events and single-channel pulses. Data from the latch is transferred in the form of bits. A detailed description of the pulse shaping and processing is presented in Appendix \ref{A1}.

%Device functionality
%Advanced trigger regime
The outputs of the latch can be directly employed for implementing real-time feed-forward control. For example, one can switch between states of an electro-optic modulator to modify the performance of the following experimental setup. The maximum input frequency is limited to 1.5~GHz by the maximum operating frequency of the delay lines. When a higher voltage is required to control the subsequent devices, it is necessary to utilize a logic-level translator. In this work, we used the microcontroller unit for controlling and monitoring overlap logic and data processing and storage. Consequently, the ECL-CMOS translator is required, reducing the maximum input frequency to 800~MHz.

%Advanced coincidence counting regime
The presented CCU allows the operation of two specific regimes depending on the complexity of the measurement. One can store the complete information about all possible coincidence events ($2^{16} = 65536$ kinds of coincidence events in total), termed as the full histogram regime. In the second regime, the CCU counts only coincidences of the same order without information about the channel number (the reduced histogram regime).
For our specific technical solution, the maximum processing rate is about 3 million events per second for the reduced histogram and 2~million events for the full histogram. This rate can be further enhanced by utilizing the FPGA.

%CCU delay lines function
Another purpose of the microcontroller unit is to configure the delay lines in the shapers to synchronize the input signals and set the length of the coincidence window. It also enables self-calibration of the CCU and monitoring of the operating temperature and other parameters. The minimum coincidence window can be set below 100~ps, and the maximum coincidence window length can reach 20~ns. The detailed electronic characterization of the CCU is presented in Appendix~\ref{A2}.
%CCU comparison
The comparison with the state-of-the-art approaches is given in Appendix~\ref{A3}.

\section{Photon statistics measurement of large optical states}
%Experimental setup
Numerous technological approaches have been developed and experimentally verified to achieve the photon-number resolution, falling into two categories: inherent photon-number resolution (superconducting nanowire single-photon detectors \cite{Cahall2017, Zhu2020, Tao2020} or transition edge sensors \cite{Lita2008, Gerrits2012, Morais2022arXiv, Eaton2022, Cheng2022, Li2023}) and multiplex/multipixel detection schemes \cite{Jex1996,Mandel1977,Braunstein2001,Hradil2003,Banaszek2003,Franson2003,Walmsley2003,TMD2004,Ramos2007,Silberhorn2007,Jezek2008,Berggren2009,Vogel2012,Bartley2013,Saleh2016,Sperling2017,Wu2017,Straka2018,Silberhorn2018,Hlousek2019,Bohmann2019}. In our case, the multiplexed detector consists of tunable beam splitters composed of a half-wave plate and a polarizing beam splitter, see Figure~\ref{fig_1}(b). This beam-splitting approach allows accurate adjustments of the splitting ratio with an absolute error below 0.3\%. Each channel is coupled to a multi-mode fiber and brought to a single-photon avalanche diode (SPAD) with efficiency ranging from 55 to 70\% at $0.8~\mu$m, 200-300~ps timing jitter, and 20-30~ns dead time. The total detection efficiency $\eta$ is defined as the ratio of the total number of detected photons to the total number of incident photons. Based on the measured transmittance of the multiport optical network and SPAD efficiencies, we experimentally determined $\eta$ of 50(1)\%. Furthermore, utilizing low-loss optics and superconducting nanowire single-photon detectors \cite{EsmaeilZadeh2021} can enhance global efficiency to over 85\%.

As a light source, we used a gain-switched semiconductor laser diode to generate a coherent nanosecond pulsed light with the central wavelength of $0.8~\mu$m. The laser diode is driven by nanosecond electronic pulses at a repetition rate of $1$~MHz. We used the temporal intensity modulation of the initial coherent light by rotating ground glass to generate the pseudo-thermal light with Bose-Einstein distribution \cite{Martienssen1964}. The optical signal is collected by a single-mode fiber to produce a single-mode thermal state.

\begin{figure}[t!] \centering
	\includegraphics[width=1.0\columnwidth]{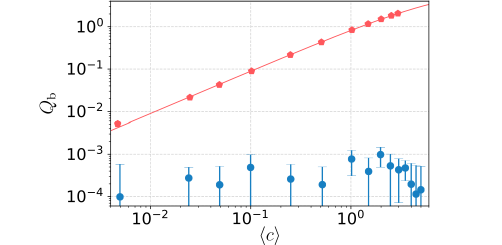}
	\caption{
		The binomial parameter $Q_{\text{b}}$ as a function of the mean number of clicks $\langle c \rangle$.
		Shown are the coherent states (blue dots) and thermal states (red pentagon); the solid line stands for the theoretical model for thermal state.
	}
	\label{fig_2}
\end{figure}

%Direct classification of light states
We quantified the statistical properties of light, focusing on coherent and thermal states, across a wide range of mean photon numbers. These states of light were detected by the ten-channel multiplexed detector and processed by the CCU in the reduced histogram regime. The coincidence window width was set to be significantly larger than both the optical pulse width and the detector jitter. 
Due to non-unity detection efficiency, noise, and a finite number of single-photon detection channels, we observe the probability distribution of the coincidence events (i.e. click statistics) $c_{m}$ instead of the photon statistics $p_{n}$. However, the click statistics still carries information about the character of the initial state of light.
The parameter quantifying this phenomenon is called the binomial parameter \cite{Vogel2012, Sperling2012} defined as $Q_{\text{b}} = \frac{\langle (\Delta c)^{2} \rangle}{\frac{\left \langle c \right \rangle}{N} \left ( 1-\frac{\left \langle c \right \rangle}{N} \right )} - 1$, where $\left \langle c \right \rangle = \sum_{i=0}^{N} ic_{i}$, $\langle (\Delta c)^{2} \rangle = \sum_{i=0}^{N} (i-\left \langle c \right \rangle)^{2}c_{i}$, and $N$ stands for number of detection channels. States of light with a Poisson distribution result in a binomial parameter of $Q_{\text{b}} = 0$, whereas non-negative values occur for super-Poissonian states (see Figure~\ref{fig_2}).
For coherent states, we measured the binomial parameters that varied between $1(5)\times10^{-4}$ and $1.0(5)\times10^{-3}$, with a mean number of clicks $\left \langle c \right \rangle$ ranging from $0.005$ to $5$.
The values of the binomial parameter do not exactly match zero due to the higher variance in the measured click statistics caused by excess noise in the light source.
Measured thermal states cover values of mean number of clicks $\left \langle c \right \rangle$ ranging from $0.002$ to $3$ with $Q_{\text{b}}$ that encompass the range from $5.1(7)\times10^{-3}$ to $2.060(1)$.
All measured values exhibit excellent agreement with the theoretical predictions.

\begin{figure}[b!] \centering
	\includegraphics[width=1.0 \columnwidth]{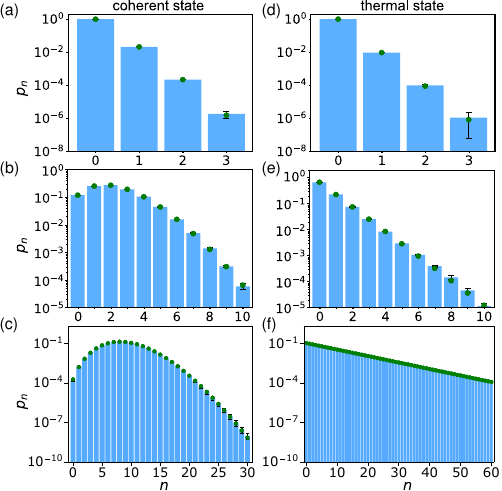}
	\caption{
		Retrieved (blue bars) and theoretical photon statistics (green dots) of coherent (left column) and thermal state (right column) for several mean photon numbers: (a) $\langle n \rangle_{\text{CS}}$ = 0.02119(3), (b) $\langle n \rangle_{\text{CS}}$ = 2.144(3), (c) $\langle n \rangle_{\text{CS}}$ = 10.21(1), (d) $\langle n \rangle_{\text{TS}}$ = 0.00949(5), (e) $\langle n \rangle_{\text{TS}}$ = 0.5113(4), and (f) $\langle n \rangle_{\text{TS}}$ = 8.41(3).
	}
	\label{fig_3}
\end{figure}

\begin{figure}[t!] \centering
	\includegraphics[width=1.0\columnwidth]{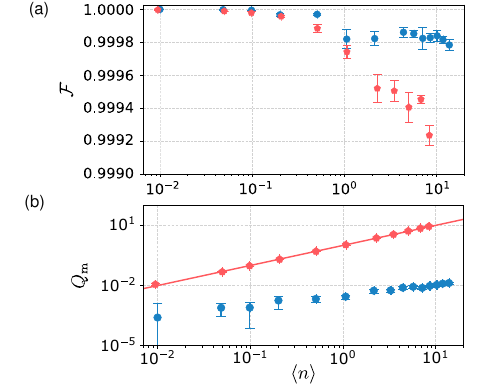}
	\caption{
		Quantification of the retrieved photon statistics $p_{n}$: (a) fidelity ${\cal F}$ and (b) Mandel parameter $Q_{\text{m}}$ as a function of mean photon number $\langle n \rangle$.
		Shown are the coherent states (blue dots) and thermal states (red pentagon); the solid line stands for the theoretical model for thermal state.
	}
	\label{fig_4}
\end{figure}

%Photon statistics characterization
In addition to directly measuring the click statistics, we have also retrieved photon statistics.
Here, we demonstrated the faithful photon statistics reconstruction over three orders of magnitude of the mean photon numbers within a dynamic range of up to 60 photons (see Figure~\ref{fig_3}).
Results show unprecedented agreement between theoretical distributions (green dots) and observed data (blue bars) down to the probabilities of $10^{-8}$.
It is important to stress here that the photon statistics retrieval includes all imperfections, such as slight imbalances of splitting ratios in the photon-number-resolving detector and imperfect light state preparation.
The fidelity of the retrieved photon statistics, defined as ${\cal F}=\textrm{Tr}[\sqrt{p_n\cdot p_n^{\textrm{\scriptsize ideal}}}]^2$, surpasses $0.999$ for all measured states of light (see Figure~\ref{fig_4}(a)).
In fact, coherent states exhibit a fidelity exceeding $0.9997$.
As the mean photon number increases, the retrieval process degrades. It is a result of the increased probability of higher Fock states within the analyzed state, leading to a progressive divergence from the detector dynamic range.
The extent of measurement degradation by the reconstruction error naturally depends on the photon statistics of the initial light state.

The presented measurement workflow is scalable with the total number of detection channels. Increasing their quantity can expand the detector dynamic range, enabling the measurement of optical states with even higher intensities. Also, the presented reconfigurable detection network allows for a decrease in the number of employed detectors to measure specific characteristics of the source under the test. Specifically, this enables a direct evaluation of the anti-correlation parameter $\alpha$ \cite{Grangier1986}, non-classicality \cite{Sperling2012, Saleh2016, Sperling2017, Bohmann2019}, and quantum non-Gaussianity \cite{Filip2011, Straka2018}.

With the knowledge of the photon statistics, we explored the Mandel parameter \cite{Mandel1977,Mandel1979,OlayaCastro2014}, which is a convenient way to characterize deviation from Poisson statistics.
The Mandel parameter is defined as the ratio of the second and first moment of photon statistic distribution $Q_{\text{m}}=\frac{\langle(\Delta n)^{2}\rangle - \langle n\rangle}{\langle n\rangle}$.
We calculated the Mandel parameter of the retrieved photon statistics as a function of the mean photon number of the initial state; see Figure~\ref{fig_4}(b).
We measured coherent states with Poisson statistics with the Mandel parameter $Q_{\text{m}}$ ranging from $3(5)\times10^{-4}$ to $1.3(2)\times10^{-2}$ and the mean photon number reaching $\langle n \rangle = 19.84(2)$.
The ideal Poisson distribution with photon number variance $\langle(\Delta n)^{2}\rangle = \langle n \rangle$ reaches the Mandel parameter $Q_{\text{m}}=0$. The measured data are influenced by excess noise, leading to a slight offset from the anticipated value of the Mandel parameter. For generated pseudo-thermal states, we obtained the Mandel parameter ranging from $Q_{\text{m}} = 0.012(2)$ to $8.93(3)$ for $\langle n \rangle$ up to $8.410(3)$.
The ideal chaotic thermal light possesses photon bunching, with the Mandel parameter $Q_{\text{m}}$ equals the mean photon number $\langle n\rangle$, as confirmed by the data (see Figure~\ref{fig_4}(b)).

\section{Conclusion}
We designed and developed a high-performance CCU, achieving low propagation delay, a well-defined adjustable coincidence window, 10~ps overall jitter, and high precision multi-coincidence counting. The CCU enables storing the complete histograms of all possible coincidence events (65536 for 16 channels) with full channel resolution or a reduced histogram carrying information only about summarized $n$-fold coincidences. Alternatively, the information about the coincidence events can be directly used as an advanced trigger to control complex photonic systems. In terms of scalability, the CCU supports extensions of the total number of input channels, easily exceeding 100. Advancements in coincidence counting and processing to a mesoscopic scale offer ground-breaking applications in quantum communications, simulations, and boson sampling machines.

The presented CCU allows for a wide range of multiple-detector experiments. We performed a photon number counting experiment employing single-photon detectors to investigate the statistical properties of the initial light. The results show unprecedented accuracy of photon statistics measurement (${\cal{F}} > 0.999$) with a dynamic range of photon-number resolution up to 60 photons. The presented measurement process is independent of the number of channels and can be scaled up to a level where hundreds of photons can be detected, which is crucial for quantum receivers, quantum tomography, photonic source benchmarking, and other applications.

\begin{acknowledgments}
This work was supported by the Czech Science Foundation (project 21-18545S). 
JH and JG also acknowledge the support by the Palacky University (JH: IGA-PrF-2022-005, JG:  IGA-PrF-2022-001 and IGA-PrF-2023-002).
\end{acknowledgments}

%Appendix A1
%%%%%%%%%%%%%%%%%%%%%%%%%%%%%%%%%%%%%building blocks%%%%%%%%%%%%%%%%%%%%%%%%%%%%%%%%%%%%%%%%%%%%%%%%%%%%
% Dual ECL Ultra-High-Speed Comparator MAX 9600  - propagation delay 500 ps
% MC100EP51 ECL D flip-flop with Reset and Differential Clock - propagation delay 350 ps
% MC100EP05 ECL 2‐Input Differential AND - propagation delay 220 ps
% MC100EP195 ECL Programmable Delay Chip - Programmable Range: 2.2 ns to 12.2 ns in 10 ps increments 
% SY89295U precision Programmable Delay line using a digital control signal, from 3.2 ns to 14.8 ns in 10 ps increments 
% STM32F429 microcontroller
%(1.65~ns + 3~ns ofset of delay line, without MCU == Dual ECL Ultra-High-Speed Comparator MAX 9600 500~ps, ECL D flip-flop with Reset and Differential Clock 350~ps, ECL Programmable Delay Chip + precision Programmable Delay line = 10~ps + ECL 2‐Input Differential AND 220~ps + ECL 2‐Input Differential AND 220~ps + ECL D flip-flop with Reset and Differential Clock 350~ps).
%%%%%%%%%%%%%%%%%%%%%%%%%%%%%%%%%%%%%%%%%%%%%%%%%%%%%%%%%%%%%%%%%%%%%%%%%%%%%%%%%%%%%%%%%%%%%%%%%%%%%%%%%
\appendix
\section{Electronic design and implementation of the coincidence counting system}
\label{A1}
\begin{figure}[t] \centering
	\includegraphics[width=1.0 \columnwidth]{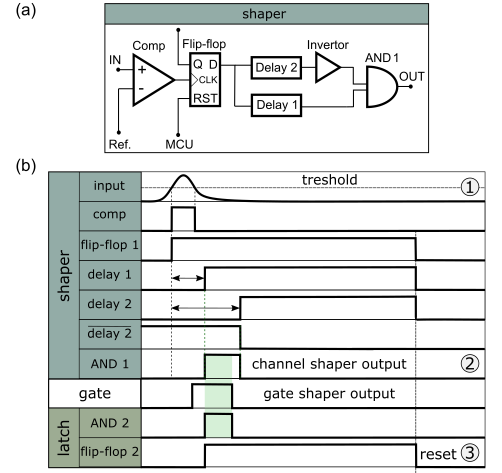}
	\caption{(a) A block diagram of a shaper, and (b) the timing diagram of the signal processing.}
	\label{fig_5}
\end{figure}

Here, we present a detailed description of the developed ultra-fast electronic multi-channel CCU based on ECL circuitry. In the main text, Figure~\ref{fig_1}(a) shows an overview of the CCU architecture consisting of a shaper, latch, and microcontroller. A shaper is a crucial circuit building block comprised of a fast comparator (MAX 9600), a pair of delay lines (SY89295U or MC100EP195, depending on market availability), and basic gates (MC100EP05). Figure~\ref{fig_5}(a) depicts a block diagram of a single shaper. For a detailed pulse shaping and processing description, see Figure~\ref{fig_5}(b). The fast comparator processes the input signal to convert it from the original waveform to a defined pulse, triggering the first flip-flop circuit. Each input channel has an adjustable threshold from 1~V to 4~V. The flip-flop stores state information and creates a time window independent of the input signal. The flip-flop output is split into two pulses and modified by parallel delay lines and invertors.

The pulse emitted by the shaper is defined as an output of the AND gate that implements the logical conjunction of the delayed pulses. The first/second delay line sets the rising/falling edge of the corresponding time window. The width of the time window is given as the time difference between these two propagation delays. The gate signal has its shaper, including one extra delay line for additional delay adjustment. The gate pulse is distributed to individual latch inputs utilizing electronic repeaters.
Recently, after acquiring data for this work, we updated the CCU design. We distribute the rising edges of the gating signal and subsequently generate individual gate pulses. This approach mitigates spreading of short edges due to long propagation on printed circuit boards and further improves the quality (steepness) of the rising and falling edges.

Each programmable delay line employed in the shaping circuitry has 1024 discrete steps with an average delay of 9~ps (for our setting and operating temperature). The resulting coincidence window is adjustable from sub-100~ps regime to approximately 20~ns, i.e. two times of the maximum delay of a single signal shaper. The delay lines can also compensate for differences in input signals arrival times. The individual channels can be delayed within a range of up to 10 ns, as discussed above. The maximum delay between the gate and the signal channels can be set to 20~ns for a sub-100~ps coincidence window. However, for the maximum 20~ns coincidence window, the entire capacity of the delay lines is used to create the coincidence window, and the input signals need to be externally synchronized.

The latch, composed of 16 AND gates and the flip-flop circuits (MC100EP51), counts all possible coincidences between the rising edge of the gate pulse and 16 rising edges of the signal pulses. The AND gates generate an output pulse only when input and gate pulses are received simultaneously. The microcontroller unit (STM32F429) receives the data from the latch and sorts the results into a histogram in memory. Data, including information about the number of single-channel pulses and coincidences, are transferred to the personal computer via USB. Finally, the microcontroller unit resets all flip-flop circuits and enables CCU for another detection event. Our CCU design offers unparalleled scalability, accommodating an unlimited number of input channels with the potential to support up to 100 input channels.

%Appendix A2
\section{Electronic characterization of the coincidence counting system}
\label{A2}
\begin{figure}[b] \centering
	\includegraphics[width=1.0\columnwidth]{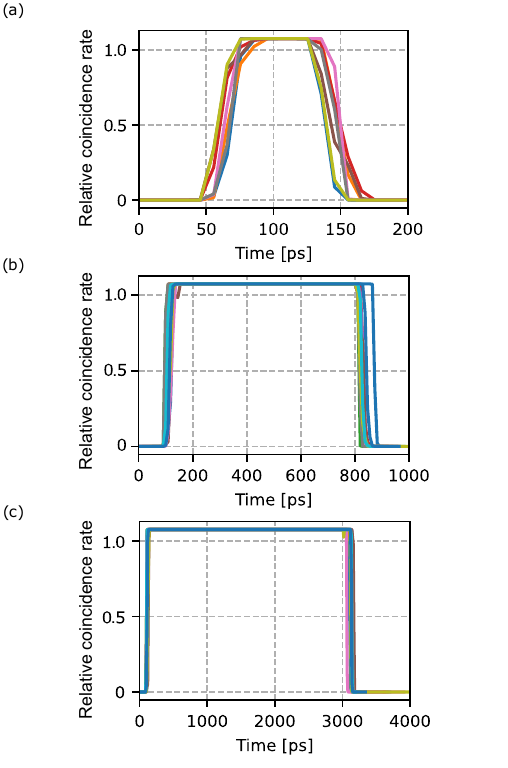}
	\caption{
		A full scan of the coincidence window of (a) sub--100~ps (b) 700~ps, and (c) 3~ns width.
	}
	\label{fig_6}
\end{figure}

\begin{figure}[h] \centering
	\includegraphics[width=1.0\columnwidth]{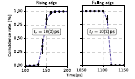}
	\caption{
		The typical achieved rising and falling edges (10 -- 90\%) of the coincidence window (width: 1 ns).
        Presented values are calculated as the average over the 250 measurement runs.
		The error bars represent one standard deviation.
	}
	\label{fig_7}
\end{figure}

To characterize the performance and capabilities of the presented device, we evaluated these figures of merit: (1)~the minimum and maximum coincidence window length, (2)~timing granularity, (3)~timing jitter, (4)~higher-order coincidence failure probability, and (5)~the maximum detection rate.
%coincidence window%
We analyzed the coincidence window width and rising and falling edges.
The whole measurement is based on changing the mutual position of the gate and channel pulse.
As a result, we got the number of coincidence events as a function of time delay.
The minimum and maximum widths of the coincidence window are sub-100~ps and 20~ns, respectively.
The main limitation of the maximum coincidence window width is the maximum range of delay lines.
The results show that the coincidence windows are well defined in time, and they are almost perfectly rectangular with sharp edges with a value typically around 18(5)~ps for all channels (see Figure~\ref{fig_6} and \ref{fig_7}).
%propagation delay
The employed high-speed ECL components with fast transition times guarantee short propagation delay below 5~ns.
%jitter
The overall jitter of the presented CCU is less than 10~ps.

%higher-order coincidence efficiency
\begin{figure}[h!] \centering
	\includegraphics[width=1.0\columnwidth]{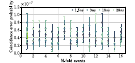}
	\caption{
		Analysis of the coincidence error probability of the individual $n$-fold coincidences for several different widths of the coincidence windows: 1.5~ns, 3.0~ns, 10~ns, and 20~ns.
		Presented values are calculated as the average over the 250 measurement runs.
		The error bars represent one standard deviation.
	}
	\label{fig_8}
\end{figure}
The presented CCU meets the conditions for a high-resolution coincidence system with equal probabilities for all orders of coincidence events. We have analyzed the precision of counting coincidences utilizing a home-built multi-channel pulse generator with a repetition rate from 0.2 to 10~MHz. The signal is generated by a relaxation oscillator whose output is delayed by a resistor–capacitor low pass network and fast inverter with Schmitt trigger inputs (74ACT14T). This generator was developed to simulate the typical output signals of the state-of-the-art single-photon detectors with a jitter of less than 10~ps. All detection channels were synchronized with the generator output pulse. The result is that all coincidence events should be evaluated as 16-fold coincidences. We compared the measured coincidence events with the number of gating pulses to evaluate the counting error of the number of coincidence events. For measurements ranging from 16-fold coincidences down to single events, we systematically deactivated the detection channels one by one. Figure~\ref{fig_8} shows the analysis of coincidence counting precision for several configurations of coincidence window width from 1.5~ns~up~to~20~ns. All coincidence errors across all tested coincidence windows are lower than $10^{-7}$. It indicates ultra-low coincidence errors due to resetting the latches, most probably caused by back reflections via imperfect impedance matching.

\section{CCU performance comparison}
\label{A3}
In Table~\ref{T1}, the performance summarization of the various CCU approaches is shown.
The presented CCU offers high n-fold coincidences counting ($\text{n}=16$) with channel-number resolution, the shortest well-defined sub--100~ps coincidence windows with 10~ps overall jitter, and the highest maximum input frequency of 1.5~GHz.
\begin{table}[h]
	\begin{tabular}{|c|c|c|c|c|c|c|c|c|}
		\hline
		Ref. & \cite{Park2015} & \cite{Jiang2016} & \cite{Gupta2018} & \cite{Jiang2018} & \cite{Arabul2020} & \cite{Park2021}  & \begin{tabular}{@{}c@{}}This \\ work\end{tabular} \\ \hline
		\begin{tabular}{@{}c@{}}Number of \\ channels\end{tabular} & 8	& 48 & 8 & 32 & 8  & 20 & 16 \\ \hline
		\begin{tabular}{@{}c@{}}Measurable \\ coinc. folds\end{tabular}	& 8 & 6 & 8& 8 & 8 & 20 & 16 \\ \hline
		\begin{tabular}{@{}c@{}}Max. input \\ frequency [MHz]\end{tabular} & 163 & 76 & 50 & 80 & 40\footnote{Up to 2-fold coincidences only} & 400\footnote{Characterized in 2-fold coincidences measurement} & 1500 \\ \hline
		\begin{tabular}{@{}c@{}}Min. coinc. \\ window [ns]\end{tabular} & 0.47 & 0.3 & 10 & 0.39 & -- & 0.46 & $<$0.1 \\ \hline
		\begin{tabular}{@{}c@{}}Max. coinc. \\ window [ns]\end{tabular} & 13.22 & 1.9 & 70 & -- & -- & 10 & 20 \\ \hline
        \begin{tabular}{@{}c@{}}Channel \\ resolution\end{tabular} & NO & NO & NO & NO & YES & NO & YES \\ \hline
	\end{tabular}
	\caption{Performance parameters comparison of the presented CCU and different coincidence unit approaches.
	}
	\label{T1}
\end{table}

\newpage
%%Bibliography
%

\end{document}